\documentclass[a4paper,twocolumn]{article}

\usepackage[latin1]{inputenc}
\usepackage{float}
\usepackage{amsmath,amssymb}
\usepackage{mathrsfs}
\usepackage{theorem}
\usepackage{graphicx}
\usepackage{color}
\usepackage{wasysym}
\usepackage{hyperref} 

\definecolor{blue}{rgb}{0,0,1}
\definecolor{green}{rgb}{0,1,0}
\definecolor{red}{rgb}{1,0,0}
\definecolor{vio}{rgb}{1,0,1}
\definecolor{ama}{rgb}{1,1,0}
\newcommand{\bc}{\begin{center}}
\newcommand{\ec}{\end{center}}
\newcommand{\be}{\nopagebreak[3]\begin{equation}}
\newcommand{\ee}{\end{equation}}
\newcommand{\ba}{\nopagebreak[3]\begin{eqnarray}}
\newcommand{\ea}{\end{eqnarray}}

{\theorembodyfont{\sffamily} \newtheorem{teorema}{Theorem}[section]}
{\theorembodyfont{\sffamily} }
{\theorembodyfont{\sffamily} }

\begin{document}

\title{\bf 
Approximation method for the
relaxed covariant form of the gravitational field equations \\
for particles
}

\author{
Emanuel Gallo
and
Osvaldo M. Moreschi
\\
{\rm \small FaMAF, Universidad Nacional de C\'{o}rdoba,}\\
{\rm \small Instituto de F\'\i{}sica Enrique Gaviola (IFEG), CONICET,}\\
{\rm \small Ciudad Universitaria,(5000) C\'{o}rdoba, Argentina.}
}




\twocolumn[
  \begin{@twocolumnfalse}
    \maketitle
    \begin{abstract}
We present a study of the so called relaxed field equations of general relativity
in terms of a decomposition of the metric; which is designed to deal with the notion of
particles.

Several known results are generalized to a coordinate free covariant discussion.

We apply our techniques to the study of a particle up to second order.
     
    \end{abstract}

 \vspace{5mm}
  \end{@twocolumnfalse}
  ]

%
%




\section{Introduction}

The notion of particle is fundamental to the Newtonian mechanics framework;
in fact, this whole theoretical framework can be constructed
in terms of the notion of test particles and massive particles.
It is then natural to ask whether this notion can be translated to other frameworks,
as is general relativity.

Within general relativity one understands  Newtonian mechanics 
as the limit of  weak field and slow motion.
So we know that one can regain the notion of particle in this regime.
Also in general relativity, the concept of test particle is a natural one,
which allows to discuss several physically interesting situations.

At first sight it is not at all clear that one can extend the notion of
particles (non-test) to the realm of general relativity.
To begin with, if one imagines a process in which one shrinks the sizes
of an object to obtain a point like object, one knows that at some
moment in the process one would end up with the formation of a black hole,
which has a characteristic size.
However, the post-Newtonian approach to compact objects is frequently
constructed in terms of the notion of particles; although post-Newtonian
systems are normally required to have weak fields and slowly moving objects.

It is interesting to note that the most simple black hole,
namely the one describing a vacuum spherically symmetric spacetime,
can be expressed in terms of the so called
Kerr-Schild decomposition. 
In this way, the Schwarzschild
black hole, whose maximal analytic extension is described in terms
of the well known causal conformal diagrams,
when expressed in the Kerr-Schild decomposition shows a
point like description in terms of the flat reference metric of the
Kerr-Schild form.

This indicates that it might be possible to give a particle notion to
a compact object in general relativity when expressed with respect to
background reference metrics.

If one intends to study the problem of a systems composed of several compact objects,
it appears as an appealing strategy to use approximation techniques for solving 
the field equations. Several problems are related to this.

In building approximation schemes for the study of the field equations in general relativity
it is often useful to recur to the relaxed form of the field equations; that we recall below.
Also, it frequently useful to decompose the physical metric in terms of a background metric.
In this work we plan to study both techniques.

In the process of decomposing the metric a key issue is the notion of gauge, since in general
one has more than one way to decompose the physical metric.
In order to study this issue we bring the techniques used by Friedrich in his study 
of the hyperbolic nature of the gravitational field equations.
We will present here a generalization of Friedrich's results that is convenient for our
discussion.

Although we work with coordinate independent expressions,
we also relate our work with the widely used harmonic gauge condition; and take the opportunity
to restate Anderson's result in a coordinate independent fashion.

An approximation scheme is suggested in which the previous studies are taking into account.

We apply our techniques to the problem of a single particle up to the second order.


\section{The decomposition of the metric}

Let us express the metric $g_{ab}$ of the spacetime $M$
in terms of a reference metric $\eta_{ab}$, such that
\begin{equation}\label{eq:getah}
g_{ab} = \eta_{ab} + h_{ab}.
\end{equation}

Let $\partial_a$ denote the torsion free metric connection
of $\eta_{ab}$ and $\nabla_a$ the torsion free metric connection
of $g_{ab}$; then one can express the covariant derivative
of an arbitrary vector $v$ by
\begin{equation}
\nabla_a v^b = \partial_a v^b + \Gamma_{a\;c}^{\;b} v^c ;
\end{equation}
and one can prove that
\begin{equation}
\Gamma_{a\;b}^{\;c}=
\frac{1}{2} g^{cd}
\left(
\partial_a h_{bd} + \partial_b h_{ad} - \partial_d h_{ab}
\right)
= \Gamma_{b\;a}^{\;c} .
\end{equation}
Let us observe that
\begin{equation}
\begin{split}
\Gamma_{a\;b}^{\;c} g_{ce} =& \Gamma_{aeb} =
\frac{1}{2} g^{cd} g_{ce}
\left(
\partial_a h_{bd} + \partial_b h_{ad} - \partial_d h_{ab}
\right) \\
=&
\frac{1}{2}
\left(
\partial_a h_{be} + \partial_b h_{ae} - \partial_e h_{ab}
\right)
.
\end{split}
\end{equation}

The relation between $\Gamma$ and the curvature tensor
can be calculated from
\begin{equation}
\begin{split}
[\nabla_a,\nabla_b] v^d =&
\left(
\partial_a \Gamma_{b\;c}^{\;d}
- \partial_b \Gamma_{a\;c}^{\;d} \right. \\
& \left. +
\Gamma_{a\;e}^{\;d} \, \Gamma_{b\;c}^{\;e}
-
\Gamma_{b\;e}^{\;d} \, \Gamma_{a\;c}^{\;e}
\right) v^c \\
& + \Theta_{abc}^{\;\;\;\;\;d} v^c \\
=&
\; R_{abc}^{\;\;\;\;\;d} v^c ;
\end{split}
\end{equation}
where $\Theta$ is the curvature of the $\partial_a$
connection.
Then
the Ricci tensor can be calculated from
\begin{equation}
\begin{split}
R_{ac} & \equiv R_{abc}^{\;\;\;\;\;b} \\
& =
\Theta_{ac} +
\partial_a \Gamma_{b\;c}^{\;b}
- \partial_b \Gamma_{a\;c}^{\;b}
+
\Gamma_{a\;e}^{\;b} \, \Gamma_{b\;c}^{\;e}
-
\Gamma_{b\;e}^{\;b} \, \Gamma_{a\;c}^{\;e} ;
\end{split}
\end{equation}
where $\Theta_{ac}$ is the Ricci tensor of the connexion
$\partial_a$.

\section{Auxiliary functions or gauge vector}
Let us consider four independent auxiliary functions $x^\mu$,
with $\mu=0,1,2,3$. Then let us observe that
\begin{equation}
g^{ab} \nabla_a \nabla_b x^\mu =
g^{ab} \nabla_a \partial_b x^\mu =
g^{ab} \partial_a \partial_b x^\mu -
g^{ab} \Gamma_{a\;b}^{\;c} \partial_c x^\mu .
\end{equation}
Then, if $I_\mu^{\;e}$ denotes the inverse of $\partial_c x^\mu$,
which exists by assumption of the independence of the set $x^\mu$,
one has
\begin{equation}
g^{ab} \Gamma_{a\;b}^{\;c} =
-
\left(
g^{ab} \nabla_a \nabla_b x^\mu -
g^{ab} \partial_a \partial_b x^\mu
\right)
I_\mu^{\;c}
=
H^\mu I_\mu^{\;c} ;
\end{equation}
where we are using
\begin{equation}
H^\mu =
-
g^{ab} \nabla_a \nabla_b x^\mu 
+
g^{ab} \partial_a \partial_b x^\mu
 . \label{eq:def-H}
\end{equation}
Alternatively, let us define the gauge vector $\mathscr{H}^c$
\begin{equation}
 \mathscr{H}^c = H^\mu I_\mu^{\;c} ;
\end{equation}
which implies
\begin{equation}
 H^\mu = \partial_c x^\mu \; \mathscr{H}^c =  \mathscr{H}(x^\mu) ;
\end{equation}
so that one also has
\begin{equation}\label{eq:H}
g^{ab} \Gamma_{a\;b}^{\;c} = \mathscr{H}^c .
\end{equation}

These equations show the relation that exist between working with a coordinate system,
given by the set of functions $x^\mu$, and the gauge vector $\mathscr{H}^c$;
which does not need any reference to coordinate systems at all.
In what follows we will try to use the covariant approach that employs the use
of the gauge vector $\mathscr{H}^c$.
We emphasize that Latin indices are abstract; and therefore our expressions are
coordinate independent and covariant.

Then, the Ricci tensor can be expressed by
\begin{equation}\label{eq:ricciH}
\begin{split}
R_{ac} &= 
\Theta_{ac}
+
\frac{1}{2}
g^{bd}
\left(
 \Theta_{bad}^{\;\;\;\;\;e} h_{ec}
+\Theta_{bcd}^{\;\;\;\;\;e} h_{ea}
+2 \Theta_{bca}^{\;\;\;\;\;e} h_{ed}
\right)
\\
&\quad +
\frac{1}{2} g^{bd} \partial_b \partial_d h_{ac}
-
\partial_{(a}
\left(
g_{c)e} \mathscr{H}^e 
\right)
+  g_{ed} \Gamma_{a\;c}^{\;e} \mathscr{H}^d
\\
&\quad -
g^{bf} g_{ed}
\Gamma_{a\;f}^{\;d}
\Gamma_{b\;c}^{\;e}
-
\frac{1}{2}
\left(
\Gamma_{a\;}^{\;bd}
\Gamma_{bcd}
+
\Gamma_{c\;}^{\;bd}
\Gamma_{bad}
\right)
.
\end{split}
\end{equation}

Let us note that if the vector field $\mathscr{H}^c$ is given by (\ref{eq:H}), then for any function
$x^\mu$ one has
\begin{equation}\label{eq:xconH}
-
g^{ab} \nabla_a \nabla_b x^\mu 
+
g^{ab} \partial_a \partial_b x^\mu = \mathscr{H}(x^\mu)
 .
\end{equation}

In the standard studies on approximations to the solution of the field equations,
one frequently finds the choice of harmonic coordinates for the set of the $x^\mu$'s;
however, in equation (\ref{eq:ricciH}) one can see that only the vector field  $\mathscr{H}^c$
appears, without any reference to a choice of auxiliary functions. Therefore one could just
refer to the gauge vector $\mathscr{H}^c$.

\section{The field equations in relaxed covariant form}

Previous to the discussion of the relaxed covariant field equations, we would like to
refer to the work of Friedrich\cite{Friedrich85} and its extension to this
coordinate independent discussion.

\subsection{Friedrich's theorem without the use of coordinates}
The field equations are
\begin{equation}
\label{eq:eins1}
R_{ac} = -8\pi \kappa
\left(
T_{ac} - \frac{1}{2}g_{ac} g^{bd} T_{bd}
\right) .
\end{equation}

Equation (3.22) in reference \cite{Friedrich85} can be obtained from (\ref{eq:ricciH})
by expressing it in a coordinate frame and neglecting the $\Theta$ terms.
In this way, one would obtain the analogous expression where all the
appearance of $\partial$ derivatives are replaced by coordinate derivatives $\partial_\mu$,
the tensors $\Gamma$ are replaced by the Chrirstoffel symbols and one uses 
$F^\epsilon = -{H}^\epsilon$;
namely:
\begin{equation}\label{eq:friedrich1}
\begin{split}
&
\frac{1}{2} g^{\beta \delta} \partial_\beta \partial_\delta g_{\alpha \sigma}
+
g_{\epsilon(\sigma}
\nabla_{\alpha)}  F^\epsilon 
\\
& 
-
g^{\beta \phi} g_{\epsilon\delta}
\Gamma_{\alpha\;\phi}^{\;\delta}
\Gamma_{\beta\;\sigma}^{\;\epsilon}
-
\frac{1}{2}
\left(
\Gamma_{\alpha\;}^{\;\beta \delta}
\Gamma_{\beta \sigma \delta}
+
\Gamma_{\sigma\;}^{\;\beta \delta}
\Gamma_{\beta\alpha \delta}
\right) \\
& = -8\pi \kappa
\left(
T_{\alpha \sigma} - \frac{1}{2}g_{\alpha \sigma} g^{\beta \delta} T_{\beta \delta}
\right)
.
\end{split}
\end{equation}

Friedrich has studied\cite{Friedrich85} this system introducing the notion of ``coordinate gauge source''
\begin{equation}
 F^\mu =  \nabla^\alpha \nabla_\alpha x^{\mu} .
\end{equation}
Subsequently, Friedrich studied the case in which $F^\mu$ is given arbitrarily.

Then, we can rephrase Friedrich's theorem in the following form:
\begin{teorema}
Let $x^\mu$ be four independent functions that are used as a coordinate system.
 If $g_{\mu\nu}$ is a solution of (\ref{eq:friedrich1}) together with the matter equations such that
on the initial surface one has $F^\mu =  \nabla^\nu \nabla_\nu x^{\mu}$,
$\nabla_\beta F^\mu =  \nabla_\beta \nabla^\nu \nabla_\nu x^{\mu}$,
then $g_{\mu\nu}$ is in fact a solution of Einstein's field equations.
\end{teorema}
This theorem can be understood in two ways. In one of them, we think that the four coordinates 
$x^\mu$ are given and then the theorem checks whether the $F^\mu$'s satisfy the above equations.
In the other way, one think that the $F^\mu$'s are given and then the theorem checks whether there exists
a coordinate system of $x^\mu$'s such that the equations in the theorem are satisfied.

From the fact that $H^\mu = -F^\mu$, one deduces, using the same techniques as in \cite{Friedrich85},
that the generalized Friedrich's theorem holds, namely, consider the four functions $H^\mu$
as given a priori, then:
\begin{teorema}
 If $g_{ab}$ is a solution of (\ref{eq:eins1}), 
with the decomposition of the metric as in (\ref{eq:getah}) and
with the Ricci tensor as given by (\ref{eq:ricciH}) with $\mathscr{H}^c = H^\mu I_\mu^{\;c}$,
 together with the matter equations such that
on the initial surface one has $H^\mu = -(g^{ab} \nabla_a \nabla_b x^\mu + g^{ab} \partial_a \partial_b x^\mu)$,
$\nabla_c H^\mu = -\nabla_c (g^{ab} \nabla_a \nabla_b x^\mu - g^{ab} \partial_a \partial_b x^\mu)$,
where $x^\mu$ are four independent scalars,
then $g_{ab}$ is in fact a solution of Einstein's field equations.
\end{teorema}

This result gives great freedom in the problem of finding solutions of the field equations
in terms of a reference metric.
Suppose that one solves (\ref{eq:eins1}) for a given vector field $H^\mu$.
Also assume that one
can solve for the functions $x^\mu$ such that $g^{ab} \nabla_a \nabla_b x^\mu = -H^\mu$.
Then, let us build a flat metric $\eta$ so that $g^{ab} \partial_a \partial_b x^\mu =0$;
which in particular can be satisfied if the $x^\mu$'s are thought as Cartesian
coordinates of $\eta$. 
In this way one would obtain $H^\mu I_\mu^e = g^{ab} \Gamma_{a\;b}^{\;e}$,
and so have a solution of the field equations.

It also might be of interest to researchers in numerical relativity, since it provides the
possibility to use any coordinate system; i.e., not necessarily an harmonic one.

Instead, one could have a proposition that does not refer to the auxiliary functions whatsoever; namely
\begin{teorema}
 If $g_{ab}$ is a solution of (\ref{eq:eins1}), 
with the decomposition of the metric as in (\ref{eq:getah}) and
with the Ricci tensor as given by (\ref{eq:ricciH}),
 together with the matter equations such that
on the initial surface one has $\mathscr{H}^c = g^{ab} \Gamma_{a\;b}^{\;c}$,
$\nabla_d \mathscr{H}^c  = \nabla_d (g^{ab} \Gamma_{a\;b}^{\;c})$,
then $g_{ab}$ is in fact a solution of Einstein's field equations.
\end{teorema}
This theorem can be understood in two ways. In one of them, we think that the metric
$\eta_{ab}$ is given and then the theorem checks whether the vector $\mathscr{H}^c$ satisfies the above equations.
In the other way, one think that the  $\mathscr{H}^c$ is given and then the theorem checks whether there exists
a metric $\eta_{ab}$ such that the equations in the theorem are satisfied.

\subsection{Relaxed covariant form of the field equations and a generalization of Friedrich's theorem}
Alternatively one can use the form of the field equations 
in terms of a slight different logic.

When we use the expression of the Ricci tensor as given by (\ref{eq:ricciH}) in 
(\ref{eq:eins1}), without assuming that $\mathscr{H}^c$ is $g^{ab} \Gamma_{a\;b}^{\;c}$, namely
\begin{equation}\label{eq:relaxed}
\begin{split}
\frac{1}{2} g^{bd} \partial_b \partial_d h_{ac}
&-
\partial_{(a}
\left(
g_{c)e} \mathscr{H}^e
\right)
+ g_{ed} \Gamma_{a\;c}^{\;d} \mathscr{H}^e
+ \Theta_{ac}
\\
&
+
\frac{1}{2}
g^{bd}
\left(
 \Theta_{bad}^{\;\;\;\;\;e} h_{ec}
+\Theta_{bcd}^{\;\;\;\;\;e} h_{ea}
+2 \Theta_{bca}^{\;\;\;\;\;e} h_{ed}
\right)
\\
& 
-
g^{bf} g_{ed}
\Gamma_{a\;f}^{\;d}
\Gamma_{b\;c}^{\;e}
-
\frac{1}{2}
\left(
\Gamma_{a\;}^{\;bd}
\Gamma_{bcd}
+
\Gamma_{c\;}^{\;bd}
\Gamma_{bad}
\right) \\
=&
-8\pi \kappa
\left(
T_{ac} - \frac{1}{2}g_{ac} g^{bd} T_{bd}
\right)
;
\end{split}
\end{equation}
we will refer to these as the \emph{relaxed field equations}\cite{Walker80},

Using the standard harmonic gauge technique, one would say:
solve the relaxed field equation in the coordinate frame, with $H^\mu = 0$, and then require the equation
\begin{equation}
\label{eq:armon1}
g^{bd} \nabla_b \nabla_d x^{\mu} =0 .
\end{equation}
In the standard approach one makes use of coordinate basis; therefore the previous statement
would be the complete story. However in our case, $H^\mu$ has a second term where two covariant derivatives 
of $x^\mu$ with respect to the metric $\eta$ appears.
At this point it is important to notice that if we have the solutions $x^\mu$ 
from (\ref{eq:armon1}) then, on constructing $\eta$
with this as a Cartesian coordinate system, one would obtain $H^\mu = 0$.

In some occasions it is preferable to work with a different set of equations.
In this respect, several authors have indicated that actually to request equation (\ref{eq:armon1}) 
is equivalent\cite{Einstein:1938yz,Anderson73,Walker80}
to demand
\begin{equation}
\label{eq:diverg1}
g^{ab} \nabla_a T_{bc} =0 .
\end{equation}
When dealing with Einstein equations in the relaxed form, and treating the vacuum case,
equation (\ref{eq:diverg1}) is understood as the condition that the divergence
of the Einstein tensor must be zero 
(which of course is identically zero in the non relaxed form).

Let us study the relation between the divergence of the energy-momentum tensor and
the vector $\mathscr{H}^e$.
One can write the relaxed field equations (\ref{eq:relaxed}) in the usual form
in which on the right hand side we have just the standard term 
$-8\pi \kappa T_{ac}$; and so on the left hand side, the terms
involving $\mathscr{H}^e$ would be
\begin{equation}
 -
\nabla_{(a}
\left(
g_{c)e} \mathscr{H}^e
\right)
+
\frac{1}{2} g_{ac} g^{ef} \nabla_e \mathscr{H}_f
;
\end{equation}
where we have used that the term $8\pi \kappa T$ contributes with the term
\begin{equation}
 - g^{ef} \nabla_e \mathscr{H}_f .
\end{equation}

Then in taking its divergence, on the left hand side, the terms
involving $\mathscr{H}^e$ are
\begin{equation}
 g^{ab} \nabla_a 
\left(
 -
\nabla_{(a}
\left(
g_{c)e} \mathscr{H}^e
\right)
+
\frac{1}{2} g_{ac} g^{ef} \nabla_e \mathscr{H}_f 
\right) .
\end{equation}
If we replace $\mathscr{H}^e$ by $\Gamma^e \equiv g^{ab} \Gamma_{a\;b}^{\;e}$,
the divergence of the left hand side would be identically zero, since the Einstein tensor
has divergence zero.
Therefore we conclude that 
the divergence of the stress energy-momentum tensor is
\begin{equation}
\begin{split}
- 8 \pi \kappa
g^{ab} \nabla_a T_{bc} =&
 g^{ab} \nabla_a 
\biggl(
 -
\nabla_{(b}
\left(
g_{c)e} (\mathscr{H}^e - \Gamma^e)
\right)
\biggr. \\
& \left. +
\frac{1}{2} g_{bc} g^{de} \nabla_d \left( g_{ef} (\mathscr{H}^f - \Gamma^f) \right)
\right) 
.
\end{split}
\end{equation}
Therefore, the stress energy-momentum is conserved if and only if
\begin{equation}
\begin{split}
 g^{ab} \nabla_a 
\biggl( &
 -
\nabla_{(b}
\left( 
g_{c)e} (\mathscr{H}^e - \Gamma^e)
\right) \biggr. \\
& \left. +
\frac{1}{2} g_{bc} g^{de} \nabla_d \left( g_{ef} (\mathscr{H}^f - \Gamma^f)\right)
\right) 
=0
.
\end{split}
\end{equation}

Working out the relations, one finds that the previous equation can be expressed as
\begin{equation}\label{eq:divt}
\begin{split}
0=
-
\frac{1}{2}
g_{ce}
 g^{ab} \nabla_a 
&
 \nabla_{b}
\left(
\mathscr{H}^e - \Gamma^e
\right)
+
\frac{1}{2}
R_{ce} (\mathscr{H}^e - \Gamma^e)
.
\end{split}
\end{equation}
Which coincides with Friedrich calculation.

It follows that if one solves equation (\ref{eq:divt}) such that on an  initial hypersurface
$\mathscr{H}^b = \Gamma^b$ and $\nabla_a \mathscr{H}^b = \nabla_a \Gamma^b$,
then the energy-momentum tensor will be conserved in the evolution of the system.
Furthermore, one also deduces that:
\begin{teorema}
 If, given the metric $\eta$, one solves the relaxed field equations for $h$
together with the matter equations,
which include the conservation of the energy-momentum tensor, 
such that $\mathscr{H}^b = \Gamma^b$ and  $\nabla_a \mathscr{H}^b = \nabla_a \Gamma^b$
on an initial hypersurface, then $g_{ab}$ is a solution of Einstein equations.
\end{teorema} 
This is a rephrasing of Friedrich's theorem applied to a decomposition of the metric
and to its general relaxed covariant form of the field equations.

It is interesting to remark that Anderson\cite{Anderson73}, using a retarded integral
expression for $h$, was able to prove the equivalence between the conservation of the energy-momentum
tensor with the harmonic gauge condition.
In relation to this let us remark that if the set of functions $x^\mu$ is obtained from the
solutions of (\ref{eq:armon1}); and one uses them as harmonic coordinates of the metric $\eta$,
then one deduces that $\Gamma^c = 0$.
And also, if $\Gamma^c = 0$, then Cartesian coordinates of $\eta$ are harmonic coordinates of $g$.
This means that we can state Anderson's result in a coordinate independent way, namely:
\begin{teorema}
 Let $h$ be the retarded solution, with respect to a flat metric $\eta$ of the relaxed field
equations together with the matter equations of state,
such that $\mathscr{H}^b = 0$, then the conservation of the energy-momentum tensor
implies that $g_{ab}$ is a solution of Einstein equations.
\end{teorema}

\section{The approximation method and the treatment of particles}

The approximation method that we introduce below, is adapted to the treatment of particles;
therefore, it is convenient to begin by treating the problem of one single particle
in the context of linearized gravity, in order to clarify some of the techniques.

\subsection{The gravitational field from one particle in linearized gravity} 

\subsubsection{The description of a particle}

Let us consider a massive point particle with mass
$m_A$ describing, in a flat space-time $(M^0,\eta_{ab})$, a curve $C$
which in a Cartesian coordinate system $x^a$ reads
\begin{equation}\label{eq:trajectoria}
 x^\mu =z^\mu(\tau),
\end{equation}
with $\tau$ meaning the proper time of the particle along $C$.

{The unit tangent vector to $C$, with respect to the flat background metric is}
\begin{equation}\label{eq:4-velocity}
{u}^\mu =\frac{dz^\mu}{d\tau},
\end{equation} 
that is, $\eta(u , u)=1$.
Now, for each point $Q=z(\tau)$ of $C$, we draw a future null cone
$\mathfrak{C}_Q$ with vertex in $Q$. If we call $x^\mu_P$
the Minkowskian coordinates of
an arbitrary
point on the cone $\mathfrak{C}_Q$, then we can define
the retarded radial distance on the null cone by
\begin{equation}\label{eq:retardedr}
r = u_\mu \left(x^\mu_P-z^\mu(\tau) \right).
\end{equation}

The energy momentum tensor $T^{(0)}_{ab}$ of a point particle
is proportional to $m v_a v_b$; where $m$ is the mass
and $v^a$ its four velocity.
{
We are distinguishing between the unit tangent vector $u^a$
and the four velocity vector $v^a$, because in future works we would like to consider the
possibility to normalize the vector $v$ with respect to a different metric.
}
In order to represent a point particle
$T^{(0)}_{ab}$ must also be proportional to a
three dimensional delta function that has support on
the world line of the particle.

We will suppose that the particle does not have multipolar structure.
Then, given an arbitrary Minkowskian frame $(x^0,x^1,x^2,x^3)$,
we will express the
energy momentum by
\begin{equation}
\begin{split}
T^{(0)ab}& (x^0=z^0(\tau_0),x^1,x^2,x^3) =
m_A v^a(\tau_0) v^b(\tau_0) \\
& \frac{\delta(x^1 - z^1(\tau_0))
\delta(x^2 - z^2(\tau_0))
\delta(x^3 - z^3(\tau_0))}
{u^0(\tau_0)}
.
\end{split}
\end{equation}

\subsubsection{The first order solution}\label{sec:firstorder}

The retarded solution, in terms of Green functions, 
for the relaxed field equations (\ref{eq:relaxed}) for particle $A$,
in which we take $\mathscr{H}^b = 0$ and $\eta$ a flat metric,
 is
\begin{equation}
h^{(1)}_{ab} = - 4 m_A\frac{v_a v_b - \frac{1}{2} \eta_{ab}}{r} ;
\end{equation}
so that in general
\begin{equation}
g^{(1)}_{ab} = \left( 1 + \frac{2 m_A}{r}\right) \eta_{ab}
- \frac{4 m_A}{r} v_a v_b
.
\end{equation}
{
In these equations we have considered the definition
\begin{equation}
 v_a \equiv \eta_{ab} v^b .
\end{equation}

}

It is interesting to realize that the exact inverse of this metric is
\begin{equation}
g^{(1)ab} =  \frac{1}{ 1 + \frac{2 m_A}{r}} \eta^{ab} 
+ \frac{\frac{4 m_A}{r}}{ 1 - \left( \frac{2 m_A}{r}\right)^2 } v^a v^b
.
\end{equation}

Note that one can solve for $h_{ab}$ for an arbitrary motion of the particle;
however, the complete solution of the problem involves having to
set also $\Gamma^b = 0$; which in terms of a coordinate frame treatment is
equivalent to the harmonic condition.
Then, recalling, as mentioned previously, that 
Anderson has proved\cite{Anderson73} the 
equivalence between the harmonic condition and the divergence free condition 
on the energy-momentum tensor;
one deduces from this, that for the case of the energy-momentum tensor of a particle
it implies its geodesic motion.

\subsection{Iterative approximation method}
Now we present a general iterative method to solve the relaxed field equations.

First of all, let us note that
given the decomposition (\ref{eq:getah}) and defining the 
tensor $\tilde h^{ab}$ from
\begin{equation}
\tilde h^{ab} = \eta^{ac} \eta^{bd} h_{cd} ,
\end{equation}
where
\begin{equation}
\eta^{ab} \eta_{bc} = \delta^{a}_{c} ,
\end{equation}
that is $\eta^{ab}$ is the inverse of $\eta_{ab}$,
one can always express the inverse $g^{ab}$ in the form
\begin{equation}
g^{ab} = \eta^{ab} - \tilde h^{ab} - d^{ab}.
\end{equation}
Then making the contraction
\begin{equation}
\begin{split}
g^{ab} g_{bc} &=
\delta^a_c -
\tilde h^{ab} h_{bc} -
d^{ab} g_{bc} =  \delta^a_c ;
\end{split}
\end{equation}
one finds
\begin{equation}
d^{ab} = -\tilde h^{ad} h_{dc} g^{cb};
\end{equation}
which can be considered an implicit equation for $d^{ab}$;
but it also shows explicitly that $d$ is quadratic in
terms of $h$.

This suggests the natural series $d_2,d_3,d_4,d_5,d_6,...$ defined by
\begin{equation}
d_2^{ab} = -\tilde h^{ad} h_{dc}
\left(
\eta^{cb}
\right) ,
\end{equation}
\begin{equation}
d_{3}^{ab} = -\tilde h^{ad} h_{dc}
\left(
\eta^{cb} - \tilde h^{cb}
\right) ;
\end{equation}
\begin{equation}
d_{n}^{ab} = -\tilde h^{ad} h_{dc}
\left(
\eta^{cb} - \tilde h^{cb} - d_{(n-2)}^{cb}
\right) ;
\end{equation}
for natural numbers $n>3$.
It is clear that $d_n$ is order $h^{(n)}$.

However, we have seen that in the first order solution for a single particle, the inverse
of the metric has a term which is conformal to the flat metric $\eta$; which it will be 
convenient to take into account. For this reason we propose the following method
of approximation where this issue is considered.

The idea is to express (\ref{eq:relaxed}) and eventually (\ref{eq:armon1})
in the form
\begin{equation}
\label{eq:retar1}
\varphi \; \eta^{ab} \partial_a \partial_b f = S(f) ;
\end{equation}
where $\varphi \; \eta^{ab}$ is the term proportional
to $\eta^{ab}$ that is contained in $g^{ab}$; while the general case would be
to consider just $\eta^{ab} \partial_a \partial_b f$ for the left hand side.
This equation can also be expressed by
\begin{equation}
\label{eq:retar1c}
\eta^{ab} \partial_a \partial_b (\varphi \;  f) =
s(\varphi \; f)+
 S(f) ;
\end{equation}
where
\begin{equation}
\label{eq:retar1d}
s(\varphi \; f) \equiv
\eta^{ab} \partial_a \partial_b (\varphi \;  f) -
\varphi \; \eta^{ab} \partial_a \partial_b f.
\end{equation}
Now one would like to solve equation (\ref{eq:retar1c}) by iterations.

Let us define the sets $f^{(j)}$ such that
for $j=0$, one takes $h=0$, $x^{\mu}$ to be harmonic
functions of the metric $\eta$ and $\varphi = 1$;
 and for $j>0$,
$f^{(j)}$ is the solution of
\begin{equation}
\label{eq:retar2}
\eta^{ab} \partial_a \partial_b (\varphi^{(j-1)}\; f^{(j)})
= s(\varphi^{(j-1)} \; f^{(j-1)})+S(f^{(j-1)}) .
\end{equation}
using the retarded Green function.
As we have seen above, $\varphi^{(1)}$ clearly arises in the first
order calculation.

The application of this method to the first order, for a single particle, 
reproduces the calculation explained in
subsection \ref{sec:firstorder}. 
Next we study this case at second order.

\subsection{The second order solution}

Let us remark that the first order solution is stationary and spherically symmetric.
This structure transports to the second order solution.

The equation for $h^{(2)}_{ab}$ is 
\begin{equation}
\label{eq:h2b}
\begin{split}
\frac{1}{2} \eta^{bd} \partial_b \partial_d h^{(2)}_{ac}
& =
\partial_{(a}
\left(
g^{(1)}_{c)e} \mathscr{H}^e
\right) + 
g^{(1)}_{ed}
\Gamma_{a\;c}^{(1)e} \mathscr{H}^d
\\
&\quad -
g^{(1)bf} g^{(1)}_{ed}
\Gamma_{a\;f}^{(1)d}
\Gamma_{b\;c}^{(1)e} \\
& \quad +
\frac{1}{2}
\left(
\Gamma_{a\;}^{(1)bd}
\Gamma_{bcd}^{(1)}
+
\Gamma_{c\;}^{(1)bd}
\Gamma_{bad}^{(1)}
\right) \\
&\quad - 8\pi
T^{(1)}_{ac}
.
\end{split}
\end{equation}

We will call the right hand side, the tensor
$\mathbb{T}$; which has the structure
\begin{equation}
\mathbb{T} = \alpha(r) dt^2 - \beta(r) \left(d\vec x \right)^2
- \gamma(r) \left(\vec x \cdot d\vec x \right)^2 ;
\end{equation}
where we are using the three dimentional notation
$\vec x = (x,y,z)$
and where
\begin{eqnarray}
\alpha &=& \frac{2 m^2}{(2 m + r)^2 (2 m - r) r}
- 8\pi m \delta_3(\vec x),\\
\beta &=& \frac{-2 m^2(4 m + r)}{(2 m + r)^2 (2 m - r) r^2} ,\\
\gamma &=& \frac{8 m^2(2 m^2 + m r - 2 r^2)}{(2 m + r)^2 (2 m - r)^2 r^4} .
\end{eqnarray}

Therefore one assumes for $h^{(2)}$
the same form, namely
\begin{equation}
h^{(2)} = A(r) dt^2 - B(r) \left(d\vec x \right)^2
- C(r) \left(\vec x \cdot d\vec x \right)^2 .
\end{equation}
In this way one has
\begin{eqnarray}
h^{(2)}_{00} &=& A , \\
h^{(2)}_{ii} &=& -(B + (x^i)^2 C), \\
h^{(2)}_{ij} &=& - x^i x^j C ;
\end{eqnarray}
where the index $i,j=1,2,3$ denote spatial coordinates.

One can see then that the equations to solve are
\begin{eqnarray}
\Box \,A &=& 2  \alpha ,\label{eq:A}\\
\Box \,C -\frac{4}{r}\frac{dC}{dr} &=& 2 \gamma ,\label{eq:C}\\
\Box \,B &=& 2(\beta + C) \label{eq:B};
\end{eqnarray}
where we are using the symbol $\Box$ to denote
$\eta^{bd} \partial_b \partial_d$.

We can solve  these equations in two ways, either using Green function techniques,
or, recalling the stationary nature of the solution, just integrating the Laplace operator.
For this presentation we choose the second option.
Let us note that for any function $f(r)$ one has that
\begin{equation}
\Box \, f(r) = - \nabla^2 f =
-\left[
\frac{1}{r^2}\frac{d}{dr}
\left(
r^2 \frac{df}{dr}
\right)
\right] .
\end{equation}

Therefore one can find $A$ by two consecutive integrations,
obtaining
\begin{equation}
\label{eq:Alaplacian}
A(r) =  \frac{1}{4 }(1 - \frac{2 m}{r})
\ln\left(\frac{1 - \frac{2 m}{r}}
{1+\frac{2 m}{r}}
\right)
+ ka_2 - \frac{ka_1}{r} .
\end{equation}
Similarly one can see that the function $C(r)$ satisfies
\begin{equation}
-\frac{1}{r^6}\frac{d}{dr}
\left(
r^6 \frac{dC}{dr}
\right) = 2 \gamma ;
\end{equation}
which after integration gives
\begin{equation}
\begin{split}
C(r)=&
\; \frac{1}{40 m^2 r^5 }
\left(  -(r^5 + 288 m^5) \ln(1 - \frac{2 m}{r}) \right. \\
& + (416  m^5 -7 r^5)\ln(1 + \frac{2 m}{r}) \\
& + 128 m^5 \ln(r)  \\
& - 8 kc_1 m^2 + 40 kc_2 m^2 r^5 - 352 m^4 r \\
& \bigg.
+ 16 m^3 r^2 - 16 m^2 r^3 + 12 m r^4
\biggr) 
 .
\end{split}
\end{equation}

Then the function $B(r)$ is given by
\begin{equation}
\begin{split}
B(r) =&
\; \frac{1}{120 m^2 r^3} \\
&\biggl( (288 m^5- 20 m^3 r^2- 30 m^2 r^3 + r^5) \biggr. \\
& \quad \ln(1 - \frac{2 m}{r})
  \\
&+
(-416  m^5 +20 m^3 r^2 -130 m^2 r^3 +7 r^5) \\
& \quad \ln(1 + \frac{2 m}{r}) - 128 m^5 \ln(r)
 \\
&
- 120 kb_1 m^2 r^2 + 120 kb_2 m^2 r^3
+ 8 kc_1 m^2 \\
&- 40 kc_2 m^2 r^5 + 352 m^4 r
 \\
& \biggl.
- 1024 m^3 r^2 - 12 m r^4
\biggr)
.
\end{split}
\end{equation}

Our choice for the integration constants is:
$ka_1 = m$, $ka_2 = 0$, $kb_1 = -\frac{181}{15} m$, 
$kb_2 = \frac{17}{15}$, $kc_1 = -\frac{74}{5} m^3$ and $kc_2 = 0$.
This choice is made taking into consideration the exact solution and
the integration of the solution coming from a Green function approach; that we will
not discuss here due to considerations of space.
By the exact functions we mean the metric components
of the Schwarzschild spacetime in harmonic coordinates;
which are
\begin{eqnarray}
A_{exact} &= & \frac{1-\frac{m}{r}}{1+\frac{m}{r}} , \\
B_{exact} &= & \left(1+\frac{m}{r}\right)^2 , \\
C_{exact} &= & \left(\frac{1+\frac{m}{r}}{1-\frac{m}{r}}\right)
     \frac{m^2}{r^4} . 
\end{eqnarray}

\begin{figure}[h!]
\centering
\includegraphics[clip,angle=-90,width=0.5\textwidth]{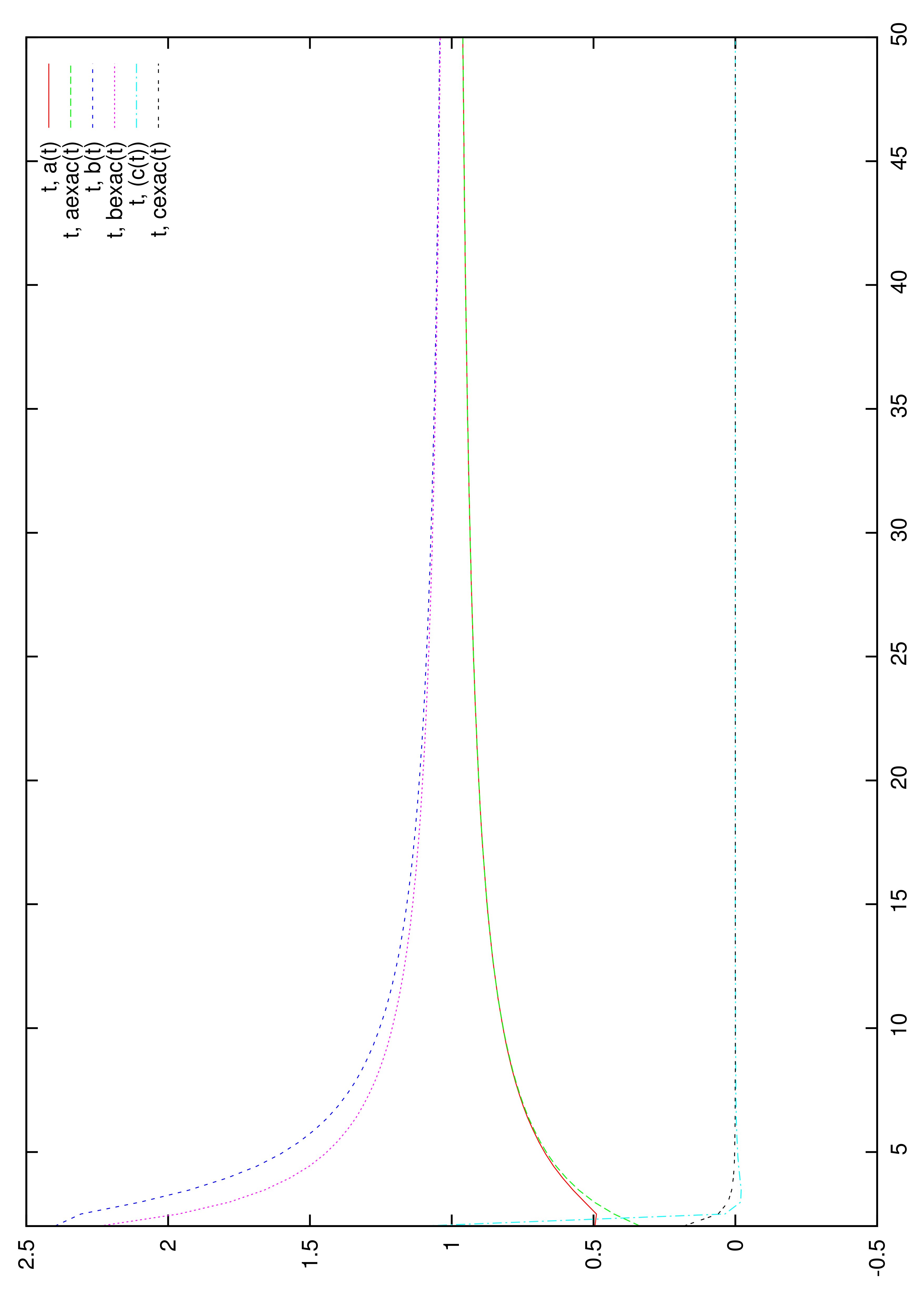}
\label{fig:comparea}
\caption{Comparison of the functions
calculated in the second iteration with the exact values.
}
\end{figure}

A graphical comparison with the exact functions of the
Schwarzschild solution in harmonic coordinates are shown in figure \ref{fig:comparea}.

It can be observed that in second order one obtains an excellent comparison of the
solution with the exact values of the metric components; even for very small values
of the radial coordinate.
Although this comparison has limited value, it is in any case remarkable that it is only
necessary to go only to second order to obtain such a good approximation.

\section{Final comments}

We have presented an study of an approach to the gravitational field equation through
the relaxed covariant form of them. The whole approach is intended to deal with 
the notion of compact objects.

The relaxed field equations was studied using Friedrich approach to the problem and
we have also refer to Anderson's result in the field of harmonic conditions.

We have generalized Friedrich results to a covariant formulation in terms of a
decomposition of the metric.

Anderson's result has been restated in a form that does not make reference to
coordinate conditions.

We have presented an approximation method that can be applied to the notion 
of particles in general relativity; and which is successful in second order
for the case of a solitary compact body.

It is our intention to apply these techniques to the problem of a binary system in 
general relativity.

\section{Acknowledgments}

We acknowledge support from CONICET and SeCyT-UNC.

%
%
%

\begin{thebibliography}{1}

\bibitem{Anderson73}
J.~L. Anderson.
\newblock {Satisfaction of deDonder and Trautman conditions by radiative
  solutions of the Einstein field equations}.
\newblock {\em Gen. Rel. Grav.}, 4:289--297, 1973.

\bibitem{Einstein:1938yz}
A.~Einstein, L.~Infeld, and B.~Hoffmann.
\newblock {The Gravitational equations and the problem of motion}.
\newblock {\em Annals Math.}, 39:65--100, 1938.

\bibitem{Friedrich85}
H.~Friedrich.
\newblock On the hyperbolicity of einstein's and other gauge field equations.
\newblock {\em Commun. Math. Phys.}, 100:525--543, 1985.

\bibitem{Walker80}
M.~{Walker} and C.~M. {Will}.
\newblock {The approximation of radiative effects in relativistic gravity -
  Gravitational radiation reaction and energy loss in nearly Newtonian
  systems}.
\newblock {\em Astrophys. J.(Letters)}, 242:L129--L133, Dec. 1980.

\end{thebibliography}

\end{document}